\documentclass[10pt,a4paper]{article}
\usepackage{amsmath}
\usepackage{amssymb}
\usepackage{amsfonts}
\usepackage{amstext}
\usepackage{amsbsy}
\usepackage[mathscr]{eucal}
\usepackage{graphicx}
\usepackage{color}
\usepackage[all]{xy}
\newcommand{\beq}{\begin{equation}}
\newcommand{\eeq}{\end{equation}}
\newcommand{\beqa}{\begin{eqnarray}}
\newcommand{\eeqa}{\end{eqnarray}}
\newcommand{\ket}[1]{| #1 \rangle}

\makeindex
\title{\Large\textbf{Segre variety, conifold, Hopf fibration, and  separable multi-qubit states}}

\author{\textit{ Hoshang Heydari}\\
        \small\textit{Institute of Quantum
Science, Nihon University,}\\
\small\textit{1-8, Kanda-Surugadai, Chiyoda-ku, Tokyo 101- 8308,
Japan }}

\date{}
\begin{document}

\maketitle \thispagestyle{empty}

\maketitle
\begin{abstract}
We establish relations between  Segre variety,  conifold, Hopf
fibration, and separable sets of pure two-qubit states. Moreover,
we investigate the geometry and topology of separable sets of pure
multi-qubit states based on a complex multi-projective Segre
variety and higher order Hopf fibration.
\end{abstract}

\section{Introduction}
Quantum entanglement\cite{Sch35,EPR35} is one of the most
interesting features of quantum theory. In quantum mechanics, the
space of pure states in is an $N+1$-dimensional Hilbert space can
be described by the complex projective space $\mathbf{CP}^{N}$.
For bipartite, pure states, the entanglement of formation can be
written in terms of concurrence \cite{Wootters98}. The connection
between concurrence and geometry is found in a map called  Segre
embedding, see D. C. Brody and L. P. Hughston \cite{Dorje99}. They
illustrate this map for a pair of qubits, and point. The Segre
embedding has also been discussed in \cite{Miyake}. There is also
another geometrical description to describe pure state called Hopf
fibration. The relation between Hopf fibration and single qubit
and two-qubit states is discussed by R. Mosseri and R. Dandoloff
\cite{Moss}. They have shown that $\mathbf{S}^{2}$ base space of a
suitably oriented $\mathbf{S}^{3}$ Hopf fibration is nothing but
the Bloch sphere, while the circular fibres represent the qubit
overall phase degree of freedom. For two-qubit states, the Hilbert
space is a seven-dimensional sphere $\mathbf{S}^{7}$, which also
allows for a second Hopf fibration which is entanglement
sensitive, with $\mathbf{S}^{3}$ fibres and a $\mathbf{S}^{4}$
base. Moreover, a generalization of Hopf fibration to three-qubit
state has been presented in Ref. \cite{Bern}, where the Hilbert
space of the three-qubit state is the fifteen-dimensional sphere
$\mathbf{S}^{15}$, which allows for the third Hopf fibration with
$\mathbf{S}^{8}$ as base and $\mathbf{S}^{7}$ as fiber. In this
paper we will describe the Segre variety, which is a quadric space
in algebraic geometry
\cite{Li2000,Musili2001,Ueno1997,Griff78,Mum76}, by giving a
complete and explicit formula for it. We will compare the Segre
variety with the concurrence of  pure, two-qubit states. The
vanishing of the concurrence of a pure two-qubit state coincides
with the Segre variety. Moreover, we will establish relations
between Segre variety, conifold and Hopf fibration.  In algebraic
geometry, a conifold is a generalization of the notion of a
manifold. Unlike manifolds, a conifold can contain conical
singularities, i.e., points whose neighborhood look like a cone
with a certain base. The base is usually a five-dimensional
manifold. Conifold are very important in string theory, i.e., in
the process of compactification of Calabi-Yau manifolds. A
Calabi-Yau manifold is a compact K\"{a}hler manifold with a
vanishing first Chern class. A Calabi-Yau manifold can also be
defined as a compact Ricci-flat K\"{a}hler manifold. Finally, we
will discuss the geometry and topology of pure multi-qubit states
based on some mathematical tools from algebraic geometry and
algebraic topology, namely the multi-projective Segre variety and
higher-order Hopf fibration. Let us start by denoting a general,
pure, composite quantum system with $m$ subsystems
$\mathcal{Q}=\mathcal{Q}^{p}_{m}(N_{1},N_{2},\ldots,N_{m})
=\mathcal{Q}_{1}\mathcal{Q}_{2}\cdots\mathcal{Q}_{m}$, consisting
of a pure state $
\ket{\Psi}=\sum^{N_{1}}_{i_{1}=1}\sum^{N_{2}}_{i_{2}=1}\cdots\sum^{N_{m}}_{i_{m}=1}
\alpha_{i_{1},i_{2},\ldots,i_{m}} \ket{i_{1},i_{2},\ldots,i_{m}} $
and corresponding Hilbert space as $
\mathcal{H}_{\mathcal{Q}}=\mathcal{H}_{\mathcal{Q}_{1}}\otimes
\mathcal{H}_{\mathcal{Q}_{2}}\otimes\cdots\otimes\mathcal{H}_{\mathcal{Q}_{m}}
$, where the dimension of the $j$th Hilbert space is given  by
$N_{j}=\dim(\mathcal{H}_{\mathcal{Q}_{j}})$. We are going to use
this notation throughout this paper, i.e., we denote a pure
two-qubit states by $\mathcal{Q}^{p}_{2}(2,2)$. Next, let
$\rho_{\mathcal{Q}}$ denotes a density operator acting on
$\mathcal{H}_{\mathcal{Q}}$. The density operator
$\rho_{\mathcal{Q}}$ is said to be fully separable, which we will
denote by $\rho^{sep}_{\mathcal{Q}}$, with respect to the Hilbert
space decomposition, if it can  be written as $
\rho^{sep}_{\mathcal{Q}}=\sum^\mathrm{N}_{k=1}p_k
\bigotimes^m_{j=1}\rho^k_{\mathcal{Q}_{j}},~\sum^N_{k=1}p_{k}=1 $
 for some positive integer $\mathrm{N}$, where $p_{k}$ are positive real
numbers and $\rho^k_{\mathcal{Q}_{j}}$ denotes a density operator
on Hilbert space $\mathcal{H}_{\mathcal{Q}_{j}}$. If
$\rho^{p}_{\mathcal{Q}}$ represents a pure state, then the quantum
system is fully separable if $\rho^{p}_{\mathcal{Q}}$ can be
written as
$\rho^{sep}_{\mathcal{Q}}=\bigotimes^m_{j=1}\rho_{\mathcal{Q}_{j}}$,
where $\rho_{\mathcal{Q}_{j}}$ is a density operator on
$\mathcal{H}_{\mathcal{Q}_{j}}$. If a state is not separable, then
it is said to be entangled state.

\section{Complex projective variety}
In this section we will review basic definition of complex
projective variety. Let $\{f_{1},f_{2},\ldots,f_{q}\}$ be
continuous functions $\mathbf{K}^{n}\longrightarrow\mathbf{K}$,
where $\mathbf{K}$ is field of real $\mathbf{R}$ or complex number
$\mathbf{C}$. Then we define real (complex) space as the set of
simultaneous zeroes of the functions
\begin{equation}
\mathcal{V}_{\mathbf{K}}(f_{1},f_{2},\ldots,f_{q})=\{(z_{1},z_{2},\ldots,z_{n})\in\mathbf{K}^{n}:
f_{i}(z_{1},z_{2},\ldots,z_{n})=0~\forall~1\leq i\leq q\}.
\end{equation}
These real (complex) spaces become a topological spaces by giving
them the induced topology from $\mathbf{K}^{n}$. Now, if all
$f_{i}$ are polynomial functions in coordinate functions, then the
real (complex) space is called a real (complex) affine variety. A
complex projective space $\mathbf{CP}^{n}$ which is defined to be
the set of lines through the origin in $\mathbf{C}^{n+1}$, that
is, $\mathbf{CP}^{n}=(\mathbf{C}^{n+1}-{0})/\sim$, where $\sim$ is
an equivalence relation define by $
(x_{1},\ldots,x_{n+1})\sim(y_{1},\ldots,y_{n+1})\Leftrightarrow\exists
\lambda\in \mathbf{C}-0$ such that  $\lambda x_{i}=y_{i} \forall
~0\leq i\leq n$. For $n=1$ we have a one dimensional complex
manifold $\mathbf{CP}^{1}$ which is very important one, since as a
real manifold it is homeomorphic to the 2-sphere $\mathbf{S}^{2}$.
Moreover every complex  compact manifold can be embedded in some
$\mathbf{CP}^{n}$. In particular, we can embed  a product of two
projective spaces into a third one. Let
$\{f_{1},f_{2},\ldots,f_{q}\}$ be a set of homogeneous polynomials
in the coordinates $\{\alpha_{1},\alpha_{2},\ldots,\alpha_{n+1}\}$
of $\mathbf{C}^{n+1}$. Then the projective variety is defined to
be the subset
\begin{equation}
\mathcal{V}(f_{1},f_{2},\ldots,f_{q})=\{[\alpha_{1},\ldots,\alpha_{n+1}]\in\mathbf{CP}^{n}:
f_{i}(\alpha_{1},\ldots,\alpha_{n+1})=0~\forall~1\leq i\leq q\}.
\end{equation}
We can view the complex affine variety
$\mathcal{V}_{\mathbf{C}}(f_{1},f_{2},\ldots,f_{q})\subset\mathbf{C}^{n+1}$
as complex cone over projective variety
$\mathcal{V}(f_{1},f_{2},\ldots,f_{q})$. We can also  view
$\mathbf{CP}^{n}$ as a quotient of the unit $2n+1$ sphere in
$\mathbf{C}^{n+1}$ under the action of $U(1)=\mathbf{S}^{1}$, that
is
$\mathbf{CP}^{n}=\mathbf{S}^{2n+1}/U(1)=\mathbf{S}^{2n+1}/\mathbf{S}^{1}$,
since every line in $\mathbf{C}^{n+1}$ intersects the unit sphere
in a circle.

\section{Hopf
fibration and two- and three-qubit states}

For a pure one-qubit state $\mathcal{Q}^{p}_{1}(2)$ with $
\ket{\Psi}=\alpha_{1}\ket{1}+\alpha_{2}\ket{2}$, where
$\alpha_{1},\alpha_{2}\in\mathbf{C}$, and
$|\alpha_{1}|^{2}+|\alpha_{2}|^{2}=1$, we can parameterize this
state as
\begin{equation}
\left(%
\begin{array}{c}
 \alpha_{1}  \\
  \alpha_{2} \\
\end{array}%
\right)=\left(%
\begin{array}{c}
  \cos(\frac{\vartheta}{2})\exp{(i(\frac{\varphi}{2}+\frac{\chi}{2}))} \\
   \cos(\frac{\vartheta}{2})\exp{(i(\frac{\varphi}{2}-\frac{\chi}{2}))}  \\
\end{array}%
\right)
\end{equation}
where $\vartheta\in[0,\pi]$, $\varphi\in[0,2\pi]$ and
$\chi\in[0,2\pi]$. The Hilbert space $\mathcal{H}_{\mathcal{Q}}$
of a single qubit is the unit 3-dimensional sphere
$\mathbf{S}^{3}\subset\mathbf{R}^{4}=\mathbf{C}^{2}$. But since
quantum mechanics is $U(1)$ projective, the projective Hilbert
space is defined up to a phase $\exp{(i\varphi)}$, so we have $
\mathbf{CP}^{1}=\mathbf{S}^{3}/U(1)=\mathbf{S}^{3}/\mathbf{S}^{1}=\mathbf{S}^{2}
$. 
Now, the first Hopf map, $ \xymatrix{\mathbf{S}^{3}
\ar[r]^{\mathbf{S}^{1}} & \mathbf{S}^{2}}$ as an $\mathbf{S}^{1}$
fibration over
 a base space $\mathbf{S}^{2}$.
For a pure two-qubit state $\mathcal{Q}^{p}_{2}(2,2)$ with
$\ket{\Psi}=\alpha_{1,1}\ket{1,1}+\alpha_{1,2}\ket{1,2}+\alpha_{2,1}\ket{2,1}+\alpha_{2,2}\ket{2,2}
$, where
$\alpha_{1,1},\alpha_{1,2},\alpha_{2,1},\alpha_{2,2}\in\mathbf{C}$
and $\sum^{2}_{k,l}|\alpha_{k,l}|^{2}=1$. The normalization
condition identifies the Hilbert space $\mathcal{H}_{\mathcal{Q}}$
to be the seven dimensional sphere
$\mathbf{S}^{7}\subset\mathbf{R}^{8}=\mathbf{C}^{4}$ and the
projective Hilbert space to be $\mathbf{CP}^{3}=
\mathbf{S}^{7}/\mathrm{U}(1).$ Thus we can parameterized the
sphere $\mathbf{S}^{7}$ as a $\mathbf{S}^{3}$ fiber over
$\mathbf{S}^{4}$, that is $ \xymatrix{\mathbf{S}^{7}
\ar[r]^{\mathbf{S}^{3}} & \mathbf{S}^{4}}$ which is called the
Hopf second fibration. This Hopf map is entanglement sensitive and
the separable states satisfy
$\alpha_{1,1}\alpha_{2,2}=\alpha_{1,2}\alpha_{2,1}$, see Ref.
\cite{Moss}.

\section{Segre variety for a general bipartite state and
concurrence}

For given general pure bipartite state
$\mathcal{Q}^{p}_{2}(N_{1},N_{2})$ we want make
$\mathbf{CP}^{N_{1}-1}\times\mathbf{CP}^{N_{2}-1}$ into a
projective variety by its Segre embedding which we construct as
follows. Let $(\alpha_{1},\alpha_{2},\ldots,\alpha_{N_{1}})$ and
$(\alpha_{1},\alpha_{2},\ldots,\alpha_{N_{2}})$ be two points
defined on $\mathbf{CP}^{N_{1}-1}$ and $\mathbf{CP}^{N_{2}-1}$,
respectively, then the Segre map
\begin{equation}\label{Segre}
\begin{array}{cc}
  \mathcal{S}_{N_{1},N_{2}}:\mathbf{CP}^{N_{1}-1}
  \times\mathbf{CP}^{N_{2}-1}&\longrightarrow \mathbf{CP}^{N_{1}N_{2}-1}\\
\end{array}
\end{equation}
\begin{equation}
\begin{array}{c}
((\alpha_{1},\ldots,\alpha_{N_{1}}),
 (\alpha_{1},\ldots,\alpha_{N_{2}}))\longmapsto
 (\alpha_{1,1},\ldots,\alpha_{1,N_{1}},\ldots,\alpha_{N_{1},1},\ldots,\alpha_{N_{1},N_{2}})\\\nonumber
\end{array}
\end{equation}
 is well defined. Next, let $\alpha_{i,j}$ be the homogeneous coordinate function
on $\mathbf{CP}^{N_{1}N_{2}-1}$. Then the image of the Segre
embedding is an intersection of a family of quadric hypersurfaces
in $\mathbf{CP}^{N_{1}N_{2}-1}$, that is
\begin{eqnarray}
\mathrm{Im}(\mathcal{S}_{N_{1},N_{2}}) &=&<\alpha_{i,k}
\alpha_{j,l}-\alpha_{i,l}  \alpha_{j,k}>= \mathcal{V}\left(
  \alpha_{i,k} \alpha_{j,l}-\alpha_{i,l}  \alpha_{j,k}
 \right).
\end{eqnarray}
This quadric space is the space of separable states and it
coincides with the definition of general concurrence
$\mathcal{C}(\mathcal{Q}^{p}_{2}(N_{1},N_{2}))$ of a pure
bipartite state \cite{Albeverio,Gerjuoy} because

\begin{eqnarray}\label{Conc}
\mathcal{C}(\mathcal{Q}^{p}_{2}(N_{1},N_{2}))&=&
\left(\mathcal{N}\sum^{N_{1}}_{j,i=1}\sum^{N_{2}}_{l,k=1}
\left|\alpha_{i,k}\alpha_{j,l}-
\alpha_{i,l}\alpha_{j,k}\right|^{2}\right)^{\frac{1}{2}},
\end{eqnarray}
where $\mathcal{N}$ is a somewhat arbitrary normalization
constant. The separable set is defined by
$\alpha_{i,k}\alpha_{j,l}= \alpha_{i,l}\alpha_{j,k}$ for all $i,j$
and $k,l$. I.e., for a two qubit state we have
$\mathcal{S}_{2,2}:\mathbf{CP}^{1}
  \times\mathbf{CP}^{1}\longrightarrow \mathbf{CP}^{3}$ and
\begin{equation}
\mathrm{Im}(\mathcal{S}_{2,2})=\mathcal{V}
\left(\alpha_{1,1}\alpha_{2,2}-\alpha_{1,2}\alpha_{2,1}\right)
\Longleftrightarrow\alpha_{1,1}\alpha_{2,2}=\alpha_{1,2}\alpha_{2,1}
\end{equation}
 is a
quadric surface in $\mathbf{CP}^{3}$ which coincides with the
space of separable set of pairs of qubits. In following section
comeback to this result.
\section{Conifold}
In this section we will give a short review of conifold. An
example of real (complex) affine variety is conifold which is
defined by
\begin{equation}
\mathcal{V}_{\mathbf{C}}(\sum^{4}_{i=1}z^{2}_{i})=\{(z_{1},z_{2},z_{3},z_{4})
\in\mathbf{C}^{4}: \sum^{4}_{i=1}z^{2}_{i}=0\}.
\end{equation}
and conifold as a real affine variety is define by
\begin{equation}
\mathcal{V}_{\mathbf{R}}(f_{1},f_{2})=\{(x_{1},\ldots,x_{4},y_{1},\ldots,y_{4})\in\mathbf{R}^{8}:
\sum^{4}_{i=1}x^{2}_{i}=\sum^{4}_{j=1}y^{2}_{j},\sum^{4}_{i=1}x_{i}y_{i}=0
\}.
\end{equation}
where $f_{1}=\sum^{4}_{i=1}(x^{2}_{i}-y^{2}_{i})$ and
$f_{2}=\sum^{4}_{i=1}x_{i}y_{i}$. This can be seen by defining
$z=x+iy$ and identifying imaginary and real part of equation
$\sum^{4}_{i=1}z^{2}_{i}=0$. As a real topological space
$\mathcal{V}_{\mathbf{R}}(f_{1},\ldots,f_{n})\subset\mathbf{R}^{n}$,
$x\in\mathcal{V}_{\mathbf{R}}(f_{1},\ldots,f_{n})$ is a smooth
point of $\mathcal{V}_{\mathbf{R}}(f_{1},\ldots,f_{n})$ if there
is a neighborhood $V$ of $x$ such that $V$ is homeomorphic to
$\mathbf{R}^{d}$ for some $d$ which is usually called the local
dimension of $\mathcal{V}_{\mathbf{R}}(f_{1},\ldots,f_{n})$ in
$x$. If there is no such neighborhood $V$, then $x$ is said to be
a singular point of
$\mathcal{V}_{\mathbf{R}}(f_{1},\ldots,f_{n})$. Now, we can call
$\mathcal{V}_{\mathbf{R}}(f_{1},\ldots,f_{n})$ a topological
manifold if all points
$x\in\mathcal{V}_{\mathbf{R}}(f_{1},\ldots,f_{n})$ are smooth.
$\mathbf{S}^{n}$ is compact, since it is a closed and bounded
subset of $\mathbf{R}^{n+1}$. Now, let us define a cone as a real
space
$\mathcal{V}_{\mathbf{R}}(f_{1},\ldots,f_{n})\subset\mathbf{R}^{n}$
with a specified point $s$ such that for all
$x\in\mathcal{V}_{\mathbf{R}}(f_{1},\ldots,f_{n})$ we have that
the line $sx\in\mathcal{V}_{\mathbf{R}}(f_{1},\ldots,f_{n})$. But
every line $s\in \mathbf{R}^{n}$ intersect any sphere $
\mathbf{S}^{n-1}$ with center $s$, the cone
$\mathcal{V}_{\mathbf{R}}(f_{1},\ldots,f_{n})$ can be determined
by a compact space
$\mathcal{B}=\mathcal{V}_{\mathbf{R}}(f_{1},\ldots,f_{n})\cap\mathbf{S}^{n-1}$
called the base space of the cone. As a real space, the conifold
is cone in $\mathbf{R}^{8}$ with top the origin and base space the
compact manifold $\mathbf{S}^{2}\times\mathbf{S}^{3}$.
  One can reformulate this relation in
term of a theorem. The conifold $
\mathcal{V}_{\mathbf{C}}(\sum^{4}_{i=1}z^{2}_{i}) $ is the complex
cone over the Segre variety $\mathbf{CP}^{1}
  \times\mathbf{CP}^{1}\simeq\mathbf{S}^{2}
  \times\mathbf{S}^{2}$. To see this let us make a complex linear
  change of coordinate $\alpha^{'}_{1,1}=z_{1}+iz_{2}$,
  $\alpha^{'}_{1,2}=-z_{4}+iz_{3}$,
  $\alpha^{'}_{2,1}=z_{4}+iz_{3}$, and
  $\alpha^{'}_{2,2}=z_{1}-iz_{2}$. Thus after this linear
  coordinate transformation we have
  \begin{equation}\label{Conifold}
    \mathcal{V}_{\mathbf{C}}(\alpha^{'}_{1,1}\alpha^{'}_{2,2}-\alpha^{'}_{1,2}\alpha^{'}_{2,1})
    =\mathcal{V}_{\mathbf{C}}(\sum^{4}_{i=1}z^{2}_{i})\subset\mathbf{C}^{4}.
\end{equation}
We will comeback to this result in section \ref{coifoldsec} where
we establish a relation between these varieties, Hopf fibration
and two-qubit state. Moreover, removal of singularity of a
conifold leads to a Segre variety which also describes the
separable two-qubit states. We will investigate this connection in
the following section. We can also define a metric on conifold as
$dS^{2}_{6}=d r^{2}+r^{2}d S^{2}_{T^{1,1}}$, where
\begin{equation}
d S^{2}_{T^{1,1}}=\frac{1}{9}\left(d\psi +\sum^{2}_{i=1}\cos
\theta_{i}d\phi_{i}\right)^{2}+\frac{1}{6}\sum^{2}_{i=1}\left(d\phi^{2}_{i}
+\sin^{2} \theta_{i}d\phi^{2}_{i}\right)^{2},
\end{equation}
is the metric on the Einstein manifold $T^{1,1}=\frac{SU(2)\times
SU(2)}{U(1)}$, with $U(1)$ being a diagonal subgroup of the
maximal torus of $SU(2)\times SU(2)$. Moreover, $T^{1,1}$ is a
$U(1)$ bundle over $\mathbf{S}^{2}\times\mathbf{S}^{2}$, where
$0\leq \psi\leq 4$ is an angular coordinate and
$(\theta_{i},\phi_{i})$ for all $i=1,2$ parameterize the two
$\mathbf{S}^{2}$, see Ref. \cite{Kleb,Morr}. One can even relate
these angular coordinate to the $\alpha^{'}_{k,l}$ for all
$k,l=1,2$ as follows $$
\begin{array}{cc}
  \alpha^{'}_{1,1}=r^{3/2} e^{\frac{i}{2}(\psi-\phi_{1}-\phi_{2})}
  \sin\frac{\theta_{1}}{2}\sin\frac{\theta_{2}}{2}
  & \alpha^{'}_{1,2}=r^{3/2} e^{\frac{i}{2}(\psi+\phi_{1}-\phi_{2})}
  \cos\frac{\theta_{1}}{2}\sin\frac{\theta_{2}}{2}  \\
  \alpha^{'}_{2,1}=r^{3/2} e^{\frac{i}{2}(\psi-\phi_{1}+\phi_{2})}
  \sin\frac{\theta_{1}}{2}\cos\frac{\theta_{2}}{2}
  & \alpha^{'}_{2,2}=r^{3/2} e^{\frac{i}{2}(\psi+\phi_{1}+\phi_{2})}
  \cos\frac{\theta_{1}}{2}\cos\frac{\theta_{2}}{2}  \\
\end{array}.
$$
Moreover, if we define the conifold as
$\mathcal{V}_{\mathbf{C}}(\sum^{4}_{i=1}z^{2}_{i})$, then we
identify the Einstein manifold $T^{1,1}$ as the intersection of
conifold with the variety
$\mathcal{V}_{\mathbf{C}}(\sum^{4}_{i=1}|z^{2}_{i}|-r^{3})$ and
$T^{1,1}$ is invariant under rotations  $SO(4)=SU(2)\times SU(2)$
of $z_{i}$ coordinate and under an overall phase rotation.
\section{Conifold, Segre variety, and a pure two-qubit state}\label{coifoldsec}
In this section we will investigate relations between  pure
two-qubit states, Segre variety, and conifold. For a pure
two-qubit state the Segre variety is given by
$\mathcal{S}_{2,2}:\mathbf{CP}^{1}
  \times\mathbf{CP}^{1}\longrightarrow \mathbf{CP}^{3}$ and
\begin{eqnarray}
\mathrm{Im}(\mathcal{S}_{2,3})&=&\mathcal{V}
\left(\alpha_{1,1}\alpha_{2,2}-\alpha_{1,2}\alpha_{2,1}\right)
\\\nonumber&=&
\mathcal{V}
(\alpha^{'2}_{1,1}+\alpha^{'2}_{2,2}+\alpha^{'2}_{1,2}+\alpha^{'2}_{2,1})\\\nonumber&=
&
 \mathbf{CP}^{1}\times\mathbf{CP}^{1}\simeq\mathbf{S}^{2}\times\mathbf{S}^{2}
 \\\nonumber&\subset&
\xymatrix{\mathbf{S}^{4} &
 \ar[l]_{\mathbf{S}^{3}}\mathbf{S}^{7}\ar[r]^-{\mathbf{S}^{1}}&
 \mathbf{S}^{7}/U(1)=\mathbf{CP}^{3}}.
\end{eqnarray}
where we have performed a coordinate transformation on  ideal of
Segre variety $\mathrm{Im}(\mathcal{S}_{2,2})$. Moreover, we have
the following commutative diagram
\[
\xymatrix{ \mathbf{S}^{7}\ar[r]^{id} \ar[d]_{\mathbf{S}^{1}} &
\mathbf{S}^{7}
\ar[d]^{\mathbf{S}^{3}} \\
\mathbf{CP}^{3}=\mathbf{S}^{7}/U(1) \ar[r]^-{\mathbf{S}^{2}} &
\mathbf{S}^{4}
  =\mathbf{S}^{7}/SU(2)=\mathbf{HP}^{1}}
\]
where $\mathbf{HP}^{1}$ denotes projective space over quaternion
number field and we have the second Hopf fibration $
\xymatrix{\mathbf{S}^{7} \ar[r]^{\mathbf{S}^{3}} &
\mathbf{S}^{4}}$. Thus we have established a direct relation
between two-qubit state, Segre variety, conic variety and Hopf
fibration. Thus the result from algebraic geometry and algebraic
topology give a unified picture of
 two-qubit state. Now, let us  investigate what
happens to our state, when we do the coordinate transformation to
establish relation between conic variety and Segre variety. By the
coordinate transformation
$\alpha^{'}_{1,1}=\alpha_{1,1}+i\alpha_{1,2}$,
  $\alpha^{'}_{1,2}=-\alpha_{2,2}+i\alpha_{2,1}$,
  $\alpha^{'}_{2,1}=\alpha_{2,2}+i\alpha_{2,1}$, and
  $\alpha^{'}_{2,2}=\alpha_{1,1}-i\alpha_{1,2}$
we perform the following map
$\ket{\Psi}=\alpha_{1,1}\ket{1,1}+\alpha_{1,2}\ket{1,2}+\alpha_{2,1}\ket{2,1}+\alpha_{2,2}\ket{2,2}
\longrightarrow\ket{\Psi^{'}}$ which is given by
\begin{eqnarray}
\ket{\Psi^{'}}&=&\alpha^{'}_{1,1}\ket{1,1}+
  \alpha^{'}_{1,2}\ket{1,2}+
  \alpha^{'}_{2,1}\ket{2,1}+
  \alpha^{'}_{2,2}\ket{2,2}\\\nonumber&=&
\alpha_{1,1}(\ket{1,1}+\ket{2,2})+i\alpha_{1,2}(\ket{1,1}-\ket{2,2})
\\\nonumber&&+i\alpha_{2,1}(\ket{1,2}+\ket{2,1})
-\alpha_{2,2}(\ket{1,2}-\ket{2,1})\\\nonumber&=&\sqrt{2}\left(
\alpha_{1,1}\ket{\Psi^{+}}+i\alpha_{1,2}\ket{\Psi^{-}}
+i\alpha_{2,1}\ket{\Phi^{+}} -\alpha_{2,2}\ket{\Phi^{-}}\right).
\end{eqnarray}
Thus the equality between Segre variety, conic variety  means that
we  rewrite a pure two-qubit state in terms of Bell's basis. For
higher dimensional space we have Segre variety but we couldn't
find any relation between these two variety.
\section{Segre variety, Hopf fibration, and multi-qubit states}
In this section, we will generalize the Segre variety to a
multi-projective space and then we will establish connections
between Segre variety for multi-qubit state and Hopf fibration. As
in the previous section, we can make
$\mathbf{CP}^{N_{1}-1}\times\mathbf{CP}^{N_{2}-1}
\times\cdots\times\mathbf{CP}^{N_{m}-1}$ into a projective variety
by its Segre embedding following almost the same procedure. Let
$(\alpha_{1},\alpha_{2},\ldots,\alpha_{N_{j}})$  be points defined
on $\mathbf{CP}^{N_{j}-1}$. Then the Segre map
\begin{equation}
\begin{array}{ccc}
  \mathcal{S}_{N_{1},\ldots,N_{m}}:\mathbf{CP}^{N_{1}-1}\times\mathbf{CP}^{N_{2}-1}
\times\cdots\times\mathbf{CP}^{N_{m}-1}&\longrightarrow&
\mathbf{CP}^{N_{1}N_{2}\cdots N_{m}-1}\\
 ((\alpha_{1},\alpha_{2},\ldots,\alpha_{N_{1}}),\ldots,
 (\alpha_{1},\alpha_{2},\ldots,\alpha_{N_{m}})) & \longmapsto&
 (\ldots,\alpha_{i_{1},i_{2},\ldots, i_{m}},\ldots). \\
\end{array}
\end{equation}
is well defined for $\alpha_{i_{1},i_{2},\ldots, i_{m}}$,$1\leq
i_{1}\leq N_{1}, 1\leq i_{2}\leq N_{2},\ldots, 1\leq i_{m}\leq
N_{m}$ as a homogeneous coordinate-function on
$\mathbf{CP}^{N_{1}N_{2}\cdots N_{m}-1}$. Now, let us consider the
composite quantum system
$\mathcal{Q}^{p}_{m}(N_{1},N_{2},\ldots,N_{m})$ and let the
coefficients of $\ket{\Psi}$, namely
$\alpha_{i_{1},i_{2},\ldots,i_{m}}$, make an array as follows
\begin{equation}
\mathcal{A}=\left(\alpha_{i_{1},i_{2},\ldots,i_{m}}\right)_{1\leq
i_{j}\leq N_{j}},
\end{equation}
for all $j=1,2,\ldots,m$. $\mathcal{A}$ can be realized as the
following set $\{(i_{1},i_{2},\ldots,i_{m}):1\leq i_{j}\leq
N_{j},\forall~j\}$, in which each point
$(i_{1},i_{2},\ldots,i_{m})$ is assigned the value
$\alpha_{i_{1},i_{2},\ldots,i_{m}}$. Then $\mathcal{A}$ and it's
realization is called an $m$-dimensional box-shape matrix of size
$N_{1}\times N_{2}\times\cdots\times N_{m}$, where we  associate
to each such matrix a sub-ring
$\mathrm{S}_{\mathcal{A}}=\mathbf{C}[\mathcal{A}]\subset\mathrm{S}$,
where $\mathrm{S}$ is a commutative ring over the complex number
field. For each $j=1,2,\ldots,m$, a two-by-two minor about the
$j$-th coordinate of $\mathcal{A}$ is given by
\begin{eqnarray}
\mathcal{C}_{k_{1},l_{1};k_{2},l_{2};\ldots;k_{m},l_{m}}&=&
\alpha_{k_{1},k_{2},\ldots,k_{m}}\alpha_{l_{1},l_{2},\ldots,l_{m}}
\\\nonumber&&-
\alpha_{k_{1},k_{2},\ldots,k_{j-1},l_{j},k_{j+1},\ldots,k_{m}}\alpha_{l_{1},l_{2},
\ldots,l_{j-1},k_{j},l_{j+1},\ldots,l_{m}}\in
\mathrm{S}_{\mathcal{A}}.
\end{eqnarray}
Then the ideal $\mathcal{I}^{m}_{\mathcal{A}}$ of
$\mathrm{S}_{\mathcal{A}}$ is generated by
$\mathcal{C}_{k_{1},l_{1};k_{2},l_{2};\ldots;k_{m},l_{m}}$  and
describes the separable states in $\mathbf{CP}^{N_{1}N_{2}\cdots
N_{m}-1}$. The image of the Segre embedding
$\mathrm{Im}(\mathcal{S}_{N_{1},N_{2},\ldots,N_{m}})$ which again
is an intersection of families of quadric hypersurfaces in
$\mathbf{CP}^{N_{1}N_{2}\cdots N_{m}-1}$ is given by
\begin{eqnarray}\label{SegreM}
\mathrm{Im}(\mathcal{S}_{N_{1},N_{2},\ldots,N_{m}})&=&<\mathcal{C}_{k_{1},l_{1};k_{2},l_{2};\ldots;k_{m},l_{m}}>\\\nonumber
&=&\mathcal{V}\left(\mathcal{C}_{k_{1},l_{1};k_{2},l_{2};\ldots;k_{m},l_{m}}\right).
\end{eqnarray}
In our paper \cite{Hosh5}, we showed that the Segre variety
defines
 the completely separable states of a general multipartite state.
  Furthermore, based on this sub-determinant, we define an entanglement measure for general
 pure bipartite  and three-partite states which coincide with generalized
 concurrence.
Let us consider a general multi-qubit state
$\mathcal{Q}^{p}_{m}(2,\ldots,2)$. For this state the Segre
variety is given by equation (\ref{SegreM}) and
\begin{eqnarray}
\mathrm{Im}(\mathcal{S}_{2,\ldots,2})&=&
\mathcal{V}\left(\mathcal{C}_{1,2;1,2;\ldots;1,2}\right)
\\\nonumber&=& \overbrace{\mathbf{CP}^{1}\times\cdots\times\mathbf{CP}^{1}}^{m ~\text{times}}
\simeq\mathbf{S}^{2}\times\cdots\times\mathbf{S}^{2}
 \\\nonumber&\subset&\xymatrix{\mathbf{S}^{2^{m}} &
 \ar[l]_{\mathbf{S}^{2^{m}-1}}\mathbf{S}^{2^{m+1}-1}\ar[r]^-{\mathbf{S}^{1}}&
 \mathbf{S}^{2^{m+1}-1}/U(1)=\mathbf{CP}^{2^{m}-1}}.
\end{eqnarray}
 We can
parameterized the sphere $\mathbf{S}^{2^{m+1}-1}$ as a
$\mathbf{S}^{2^{m}-1}$ fiber over $\mathbf{S}^{2^{m}}$, that is $
\xymatrix{\mathbf{S}^{2^{m+1}-1} \ar[r]^-{\mathbf{S}^{2^{m}-1}} &
\mathbf{S}^{2^{m}}}$ which are higher order Hopf fibration.
Moreover, we have the following commutative diagram
\[
\xymatrix{ \mathbf{S}^{2^{m+1}-1}\ar[r]^{id}
\ar[d]_{\mathbf{S}^{1}} & \mathbf{S}^{2^{m+1}-1}
\ar[d]^-{\mathbf{S}^{2^{m}-1}} \\
\mathbf{CP}^{2^{m}-1}=\mathbf{S}^{2^{m+1}-1}/U(1)
\ar[r]^-{\mathbf{S}^{2^{m}-2}} & \mathbf{S}^{2^{m}}}
\]
Thus we have established relations between Segre variety and
higher order Hopf fibration and separable set of a multi-qubit
state. As an example, let us look at a pure three-qubit state. For
such state we have
\begin{eqnarray}
\mathrm{Im}(\mathcal{S}_{2,2,2})&=&
\mathcal{V}\left(\mathcal{C}_{1,2;1,2;1,2}\right)
\\\nonumber&=&
\langle\alpha_{1,1,1}\alpha_{2,1,2}-\alpha_{1,1,2}\alpha_{2,1,1},
\alpha_{1,1,1}\alpha_{2,2,1}-\alpha_{1,2,1}\alpha_{2,1,1}\\\nonumber&&
,\alpha_{1,1,1}\alpha_{2,2,2}-\alpha_{1,2,2}\alpha_{2,1,1},
\alpha_{1,1,2}\alpha_{2,2,1}-\alpha_{1,2,1}\alpha_{2,1,2}
\\\nonumber&&
,\alpha_{1,1,2}\alpha_{2,2,2}-\alpha_{1,2,2}\alpha_{2,1,2},
\alpha_{1,2,1}\alpha_{2,2,2}-\alpha_{1,2,2}\alpha_{2,2,1}
\\\nonumber&&
,\alpha_{1,1,1}\alpha_{1,2,2}-\alpha_{1,1,2}\alpha_{1,2,1},
\alpha_{1,1,1}\alpha_{2,2,2}-\alpha_{1,2,1}\alpha_{2,1,2}\\\nonumber&&
,\alpha_{1,1,2}\alpha_{2,2,1}-\alpha_{1,2,2}\alpha_{2,1,1}, ,
\alpha_{2,1,1}\alpha_{2,2,2}-\alpha_{2,1,2}\alpha_{2,2,1}
\\\nonumber&&
,\alpha_{1,1,1}\alpha_{2,2,2}-\alpha_{1,1,2}\alpha_{2,2,1}
,\alpha_{1,2,1}\alpha_{2,1,2}-\alpha_{1,2,2}\alpha_{2,1,1} \rangle
\\\nonumber&=&\mathbf{CP}^{1}\times\mathbf{CP}^{1}\times\mathbf{CP}^{1}
\simeq\mathbf{S}^{2}\times\mathbf{S}^{2}\times\mathbf{S}^{2}\\\nonumber&\subset&
\xymatrix{\mathbf{S}^{8} &
 \ar[l]_-{\mathbf{S}^{7}}\mathbf{S}^{15}\ar[r]^-{\mathbf{S}^{1}}&
 \mathbf{S}^{15}/U(1)=\mathbf{CP}^{7}}.
\end{eqnarray}
This is what we have expected to see. Moreover, we have the
following commutative diagram
\[
\xymatrix{ \mathbf{S}^{15}\ar[r]^{id} \ar[d]_{\mathbf{S}^{1}} &
\mathbf{S}^{15}
\ar[d]^{\mathbf{S}^{7}} \\
\mathbf{CP}^{7}=\mathbf{S}^{15}/U(1) \ar[r]^-{\mathbf{S}^{6}} &
\mathbf{S}^{8}}
\]
where we have the third Hopf fibration $ \xymatrix{\mathbf{S}^{15}
\ar[r]^{\mathbf{S}^{7}} & \mathbf{S}^{8}}$ for three-qubit state
which has been discussed in Ref. \cite{Bern}.
\section{Conclusion}

In this paper, we  have discussed a geometric picture of the
separable  pure two-qubit states
 based on Segre variety, conifold, and Hopf fibration. We have
 shown that these varieties and mappings give a unified picture of
 two-qubit states. Moreover, we have discussed the geometry and topology of pure
multi-qubit states based on multi-projective Segre variety and
higher-order Hopf fibration.  Thus we have established relations
between algebraic geometry, algebraic topology and
 fundamental quantum theory of entanglement. Perhaps, these
 geometrical and topological visualization puts entanglement in a broader perspective and
 hopefully gives some hint about how we can solve the problem of
 quantify entanglement.

\begin{flushleft}
\textbf{Acknowledgments:} This work was supported by the
Wenner-Gren Foundation
\end{flushleft}


\end{document}